\documentclass[runningheads]{llncs}
\usepackage{amsmath}
\usepackage{amssymb}
\usepackage[T1]{fontenc}
\usepackage{marvosym}
\usepackage{graphicx}
\begin{document}
\title{Graph Neural Networks in EEG-based Emotion Recognition: A Survey}
%
%
\newcommand{\corr}{\unskip$^{(\mbox{\small\Letter})}$}

\author{Chenyu Liu\inst{1}\thanks{Equal contribution.} \and
Yuqiu Deng\inst{2}\unskip$^{\star}$ \and
Yihao Wu\inst{1} \and
Ruizhi Yang\inst{3} \and
Zhongruo Wang\inst{4} \and
Liangwei Zhang\inst{5} \and
Siyun Chen\inst{1} \and
Tianyi Zhang\inst{1,6} \and
Yang Liu\inst{1} \and
Yi Ding\inst{1} \and \\
Liming Zhai\inst{7}\corr \and
Ziyu Jia\inst{8}\corr \and
Xinliang Zhou\inst{1}}
\authorrunning{C. Liu et al.}
\institute{Nanyang Technological University, Singapore \and
Xi’an Jiaotong University, Xi’an, China \and
Imperial College London, London, UK \and
Amazon, Seattle, WA, USA \and
Carnegie Mellon University, Pittsburgh, PA, USA \and
Uber Technologies, Inc., USA \and
School of Computer Science, Central China Normal University, Wuhan, China\\
\email{limingzhai@ccnu.edu.cn} \and
Institute of Automation, Chinese Academy of Sciences, Beijing, China\\
\email{jia.ziyu@outlook.com}}
%
%
\maketitle              
\begin{abstract}
    Compared to other modalities, EEG-based emotion recognition can intuitively respond to the emotional patterns in the human brain and, therefore, has become one of the most concerning tasks in the brain-computer interfaces. Since dependencies within brain regions are closely related to emotion, a significant trend is to develop Graph Neural Networks (GNNs) for EEG-based emotion recognition. However, brain region dependencies in emotional EEG have physiological bases that distinguish GNNs in this field from those in other time series fields. Besides, there is neither a comprehensive review nor guidance for constructing GNNs in EEG-based emotion recognition. In the survey, our categorization reveals the commonalities and differences of existing approaches under a unified framework of graph construction. We analyze and categorize methods from three stages in the framework to provide clear guidance on constructing GNNs in EEG-based emotion recognition. In addition, we discuss several open challenges and future directions, such as temporal fully-connected graph and graph condensation.
\end{abstract}
\section{Introduction}

Emotion is an integral and complex aspect of human cognition that is decisive in human decision-making, behavior, and social interaction. Therefore, Emotion Recognition is essential in areas such as the diagnosis of mental disorders and Brain-Computer Interfaces (BCIs). Human emotions have various external manifestations, such as body language, voice, expression, and physiological signals. Among them, the Electroencephalogram (EEG) signals are perceived as a product of coordinated neural activity and electrical signal transmission within the human brain, which can directly and objectively reveal genuine human emotion compared to other manifestations and thus are known as one of the most unique and vital data in emotion recognition. EEG records emotional activity from different brain regions and, therefore, can reflect the complex interactions between brain regions in emotional states and the strong correlation between specific brain regions and specific emotions \cite{min2022emotion}. In summary, inferring and exploiting complex dependencies between brain regions in emotional EEG has become a significant research direction in EEG-based emotion recognition.


Graph Neural Networks (GNNs) emerge as a powerful tool for modeling dependencies of emotional EEG within the network neuroscience framework. As networks that manipulate graph-structured data, GNNs can effectively extract features utilizing dependencies between brain regions in emotional EEG. These dependencies represent the connectivity patterns and interactions of brain regions in emotional states, which are directly associated with specific emotional activities. Therefore, benefiting from their dependencies inference ability, GNNs designed for EEG-based emotion recognition can improve classification tasks compared to traditional analysis methods and potentially uncover new insights in neuroscience. There has been a significant growth in the number and proportion of methods using GNNs in the EEG-based emotion recognition field over the last six years.

Motivated by the increasing number of recent papers proposing GNNs for EEG-based emotion recognition, and the design of these GNNs differs from methods in other time-series tasks, there is an urgent need for a comprehensive review of the GNNs in this field. First, there is currently no unified treatment for GNN structure in the EEG-based emotion recognition field. Second, brain region dependencies in emotional states have specific physiological bases, distinguishing GNNs in EEG-based emotion recognition from those used in other time-series fields.
Therefore, we categorize existing GNNs in EEG-based emotion recognition based on stages within a unified GNN construction framework. Specifically, this categorization splits the existing methods regarding node-level, edge-level, and graph-level stages and proposes a taxonomic view for each stage. In summary, the contributions of this survey are summarized below:

\begin{itemize}
  \item This survey provides a comprehensive and systematic review of existing GNNs in EEG-based emotion recognition. To the best of our knowledge, this is the first and only survey work on such a topic.
  \item We propose a novel categorization of existing GNNs in EEG-based emotion recognition, which provides clear guidance for constructing specific GNNs according to a unified framework.
  \item We summarize and highlight future directions to facilitate GNN-based works in EEG-based emotion recognition.
\end{itemize}

\begin{figure}[t]
\centering
\includegraphics[width=0.9\columnwidth]{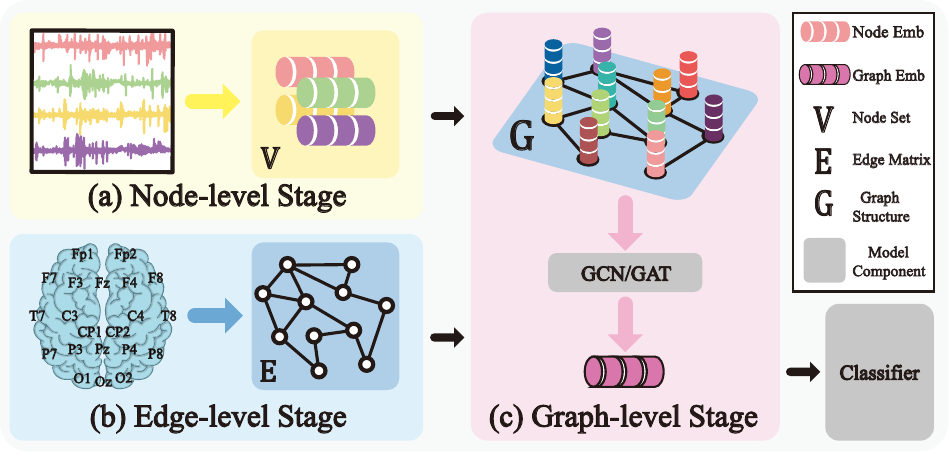}
\vspace{-10pt}
\caption{Unified framework of EEG-based emotion recognition. First, (a) indicates the selection of node features. Then, (b) refers to the computation of the edge matrix representing brain connectivity patterns. Finally, (c) denotes the construction of a graph, which differs across methods.}
\centering
\label{fig: frame}
\end{figure}

\section{Overview of Categorization}

In recent years, EEG-based emotion recognition has seen an influx of methods with GNNs. These methods focus on different aspects of designing GNNs to infer dependencies within emotional EEG. They generally answer these three critical questions during the graph construction process: What features are chosen as nodes? How to calculate edges? Which graph structure is utilized? Therefore, categorizing these methods at the model level is neither instructive nor sufficiently in-depth. To better understand these methods and propose a construction guideline for GNNs in EEG-based emotion recognition, our categorization is based on a unified framework with its corresponding stages. 

\subsection{Framework for specific GNNs}

EEG-based emotion recognition task for GNNs can be indicated as taking emotional EEG $X \in \mathbb{R}^{C \times S}$ and corresponding label $Y \in \mathbb{R}^{1}$ as input, constructing graphs and predicting emotional labels $Y' \in \mathbb{R}^{1}$ via graph embeddings. 
Let $\textbf{G}(\textbf{V},\textbf{E})$ denote a graph. $\textbf{V} \in \mathbb{R}^{N_v \times D}$ represents the node set, where $N_v = C$ indicates the number of channels and $D$ denotes node embedding dimension. $\textbf{E} \in \mathbb{R}^{N_e}$ represents the edge matrix, where $N_e = C^2$ denotes the number of edges. The goal of the model can be represented as learning a function $f(\textbf{G}) \rightarrow Y'$ that maps the graph to the corresponding emotional label. 
Therefore, the unified framework for building GNNs in EEG-based emotion recognition consists of three stages, as shown in Fig. \ref{fig: frame}.
First, the Node-level Stage represents the model selection of features to be used as nodes $\textbf{V}$. Second, the Edge-level Stage indicates that the model computes the edge matrix $\textbf{E}$. Finally, the Graph-level Stage denotes the process of graph $\textbf{G}$ construction.

\subsection{Stages of Framework}
\vspace{-10pt}
\begin{figure*}[htbp]
\centering
\includegraphics[width=1\columnwidth]{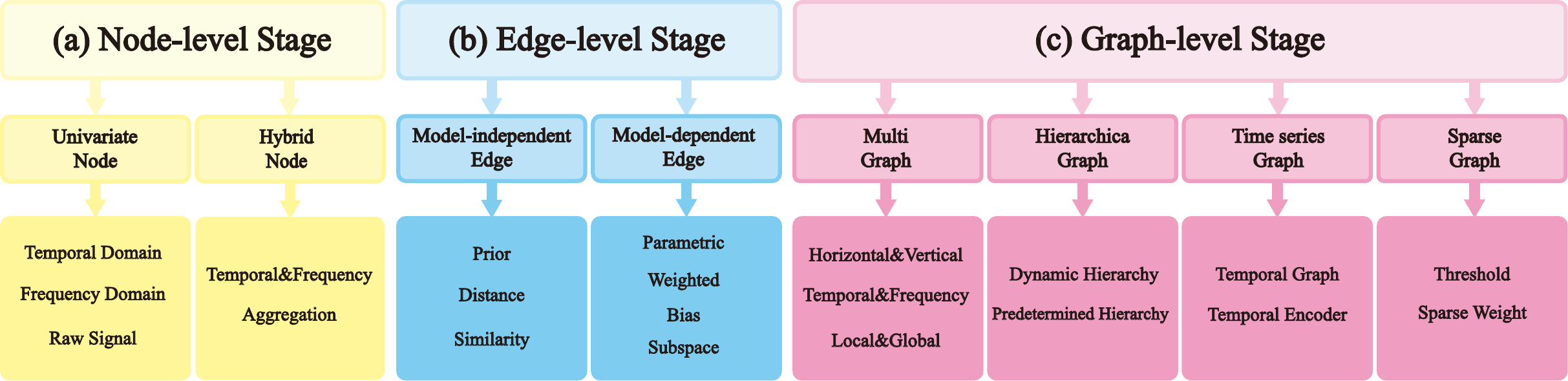}
\vspace{-10pt}
\caption{An overview of the categorization. The existing method was split into three stages for further categorization. These three steps correspond to (a) the selection of node features, (b) the calculation of edge matrices, and (c) the construction of graphs with different structures.}
\centering
\label{fig: main}
\end{figure*}

\textbf{Node-level Stage} indicates the process of selecting features as nodes $\textbf{V}$. Instead of focusing on the detailed composition and parameters of the node embedding extractor, this survey focuses on the composition of node features. Therefore, we categorize this stage according to the type of node features, as shown in Fig. \ref{fig: main} (a). These two categories are Univariate node, which utilizes a single feature, and Hybrid node, which employs multiple features. The details of each category will be explained in Section \ref{sec: node}.

\noindent \textbf{Edge-level Stage} represents the calculation of the edge matrix $\textbf{E}$. The edge matrix represents the relationships between electrode nodes. Matrices representing such relationships differ in different methods, such as the adjacency matrix or the Laplacian matrix. We uniformly regard them as edge matrices. Depending on whether the model parameters are involved in the calculation, we categorize the Edge-level Stage of the existing methods into two categories: Model-independent edge and Model-dependent edge, as shown in Fig. \ref{fig: main} (b). The sub-categories of each category will be explained in detail in Section \ref{sec: edge}.

\noindent \textbf{Graph-level Stage} denotes the process of modeling different types of graph structures $\textbf{G}$. 
The basic units utilized by existing methods for reasoning about graph embedding are Graph Convolutional Networks (GCNs) or Graph Attention Networks (GATs). The perspective of this survey is not on the detailed model components but on the graph structure used to represent the dependencies within emotional EEG. The graph structure of existing methods can be categorized into four types, as shown in Fig. \ref{fig: main} (c). Multi-graph indicates that the model employs a multi-stream structure to simultaneously construct different graphs to incorporate multiple dependencies of emotional EEG. Hierarchical graph denotes that the model organizes nodes into multiple groups to infer dependency between brain regions related to emotion. 
Time series graph is a series of graphs in the temporal dimension that represent the temporal dependency of emotional EEG. Sparse graph structure represents sparsely connected graphs that match the concentration of brain activity in emotional states. The sub-categories of each category will be explained in detail in Section \ref{sec: graph}.

\section{Node-level Stage}
\label{sec: node}
Depending on the node feature, two categories are included in the Node-level Stage: Univariate node and Hybrid node. Univariate node is further categorized into Temporal domain node, Frequency domain node, and raw signal node. Hybrid node contains Temporal\&Frequency node and Aggregation node.

\subsection{Univariate node} 
Univariate nodes refer to the most common way of using a single feature as nodes. The vast majority of existing methods utilize this method. It contains three categories, as shown in Table \ref{tb: univ}.

\noindent \textbf{Temporal domain} node implies that the model employs the temporal domain feature as nodes. 
This is the most commonly used node in the current methods, especially the differential entropy (DE) feature. 
The popularity of the DE feature, in addition to the fact that many open-source EEG-based emotion recognition datasets include this feature, such as SEED \cite{zheng2015investigating}, DE feature itself is suitable for measuring the uncertainty or randomness of the signal, which is adequate for describing the diverse dependencies in the emotional EEG. In addition to this, HD-GCN \cite{hd-gcn} and SparseDGCNN \cite{sparsedgcnn} employ the Differential Asymmetry of Synchronisation for Multivariate Signals (DASM) feature.

\noindent \textbf{Frequency Domain} node indicates that utilizing frequency domain features as nodes. The most commonly used is the Power Spectral Density (PSD) feature obtained by Fourier Transform to describe the power distribution of the signal in the frequency domain. From comparing the experiment results of existing methods, the performance of using Frequency domain node is generally slightly worse than using Temporal domain node.

\noindent \textbf{Raw Signal} node represents directly using the raw emotional EEG as nodes. This node is the least common in EEG-based emotion recognition tasks because raw emotional EEG is highly individualized, which retains more emotion-independent disturbances. Besides, the dimension of Raw signal node is higher compared to other nodes. However, this node maximizes information retention and enables end-to-end training of the model.

\begin{table}[t]
\caption{Overview of Univariate node.}
\vspace{-8pt}
\centering
\renewcommand{\arraystretch}{1.0}
\setlength{\tabcolsep}{6pt}
\resizebox{0.95\columnwidth}{!}
{
\scriptsize
\begin{tabular}{c|c|c}
\hline \hline
\multicolumn{1}{c|}{\rule[-1ex]{0pt}{3.5ex}\footnotesize \textbf{Node}} &
\multicolumn{1}{c|}{\footnotesize \textbf{Baseline}} &
\multicolumn{1}{c}{\footnotesize \textbf{Node Feature}} \\ \hline \hline

Temporal domain &
\begin{tabular}[c]{@{}c@{}}
All except the below
\end{tabular} &
\begin{tabular}[c]{@{}c@{}}
Mainly Differential Entropy
\end{tabular} \\ \hline

Frequency domain &
\begin{tabular}[c]{@{}c@{}}
LR-GCN \cite{lr-gcn}\\
GECNN \cite{gecnn} \\
TARDGCN \cite{tardgcn}\\
OMHGL \cite{omhgl} \\
GIGN \cite{gign} \\
EmT \cite{ding2025emt}
\end{tabular} &
\begin{tabular}[c]{@{}c@{}}
Power Spectral Density \\
Hilbert-Huang transform spectrum \\
Welch \\
Power Spectral Density \\
Power Spectral Density \\
Power Spectral Density
\end{tabular} \\ \hline

Raw signal &
\begin{tabular}[c]{@{}c@{}}
HetEmotionNet \cite{hetemotionnet}\\
LGGNet \cite{lggnet} \\
RPGCN \cite{zhou2025rpgcn} \\
VSGT \cite{liu2024vsgt} \\
VBH-GNN \cite{liu2024vbh}
\end{tabular} &
\begin{tabular}[c]{@{}c@{}}
Time-domain Amplitude \\
Power Conversion \\
Linear Embedding \\
Time-domain Amplitude\\
Time-domain Amplitude
\end{tabular} \\ \hline \hline

\end{tabular}%
}
\label{tb: univ}
\end{table}

\subsection{Hybrid node} 
Hybrid node implies that the node mixes a variety of features. It is consistent with the goal of Multi-graph mentioned in Section \ref{sec: multi}, which is inferring multiple dependencies within emotional EEG. 
The difference is that Multi-graph builds multiple graphs for different dependencies, while Hybrid nodes fuse features that contain different dependencies. Hybrid nodes can utilize the complementarity between different information to provide a more comprehensive view of emotional EEG dependencies. However, it increases the dimension of the node. Besides, due to the different distributions of features, it can easily lead to overfitting when the amount of data is insufficient. Hybrid node contains two categorizations.
CR-GAT \cite{cr-gat}, DBGC-ATFFNet-AFTL \cite{dbgc-atffnet-aftl}, and EGFG \cite{egfg} employ \textbf{Temporal\&Frequency} node that fuses DE features, and PSD features to infer both temporal and frequency dependencies simultaneously. Methods like ERHGCN \cite{erhgcn} utilize \textbf{Aggregation} node that aggregates a variety of features, including DASM, Average Size of Messages (ASM), etc., to preserve as many dependencies as possible.

\section{Edge-level Stage}
\label{sec: edge}

Edge-level Stage contains Model-independent edge and Model-dependent edge. Model-independent edge is divided into Prior edge, Distance edge, and Similarity edge according to the calculation of edges. Model-dependent edge is divided into Parametric edge, Weighted edge, Bias edge, and Subspace edge depending on the usage of parameters.

\subsection{Model-independent edge}
Model-independent edge implies that the calculation of the edge does not include the model parameters and backpropagation does not update this edge.
It is based on existing brain physiology paradigms or signal processing paradigms.
The fact that such edges are independent of the data distribution and model and conform to physiological or signaling patterns increases the plausibility and stability of Model-independent edge.
Moreover, such an edge does not increase model parameters and is compatible with any model structure. 
The Model-independent edge contains three edges, as shown in Table \ref{tb: model-inde}

\noindent \textbf{Prior} edge refers to the edge that is set manually based on prior knowledge. Prior Knowledge refers to existing physiological paradigms about electrode connections, which is physiologically plausible.
However, emotional EEG has individual differences, so the connections between electrodes differ between subjects. This makes the rationality of Prior edge limited because it cannot be automatically updated. Based on prior knowledge that the strength of connections between brain regions decays as a function of the inverse square of the physical distance \cite{salvador2005neurophysiological}, RGNN \cite{rgnn}, and DAGAM \cite{dagam} use 3D co-ordinates to calculate the physical distances between electrodes as the edges. The International 10-20 EEG Electrode Standard depicts the connectivity between electrodes, and MD-GCN \cite{md-gcn} and SFE-Net \cite{sfe-net} transform it into an edge matrix. LGGNet \cite{lggnet} limits the connectivity between electrodes in specific brain regions based on the distribution of brain regions.

\noindent \textbf{Distance} edge refers to treating signals as vectors and taking vector distances as edges. Calculating the distance between vectors is commonly used to measure the relationship between signals. Distance edge can be updated as the node changes, but this relationship is physiologically independent of emotion. TARDGCN \cite{tardgcn} and OMHGL \cite{omhgl} apply Cosine distance between nodes as edges. Similarly, EGFG \cite{egfg} and JSCFE \cite{jscfe} utilize Euclidean distance. MST-GNN \cite{mst-gnn} uses the Phase Lag Index, which analyses the phase distance between node signals.

\noindent \textbf{Similarity} edge refers to calculating the similarity of signals as edges.
It differs from Distance edge only in the mathematical meaning, so Similarity edge is dynamic but does not always conform to the physiological paradigm. 
MD-AGCN \cite{md-agcn} and ST-GCLSTM \cite{st-gclstm} utilize the Pearson Correlation Coefficient to compute the strength of the linear relationship between signals as their edge.
HB-SR \cite{hb-sr} uses Phase Locking Value to calculate the degree of phase similarity between electrode signals at a particular frequency as their edge.
In addition, HD-GCN \cite{hd-gcn} applies the Gaussian similarity, and HetEmotionNet \cite{hetemotionnet} indicates mutual information to represent the correlation between two signals as the edge.

\begin{table}[t]
\caption{Overview of Model-independent edge.}
\vspace{-10pt}
\centering
\renewcommand{\arraystretch}{1.0} 
\setlength{\tabcolsep}{10pt}
\resizebox{0.95\columnwidth}{!}
{
\scriptsize
\begin{tabular}{c|c|c}
\hline \hline
\multicolumn{1}{c|}{\rule[-1ex]{0pt}{3.5ex}\footnotesize \textbf{Edge}} &
\multicolumn{1}{c|}{\footnotesize \textbf{Baseline}} &
\multicolumn{1}{c}{\footnotesize \textbf{Edge Feature}} \\ \hline \hline

Prior &
\begin{tabular}[c]{@{}c@{}}
RGNN \cite{rgnn} \\
SFE-Net \cite{sfe-net} \\
MD-GCN \cite{md-gcn} \\
LGGNet \cite{lggnet} \\
DAGAM \cite{dagam}
\end{tabular} &
\begin{tabular}[c]{@{}c@{}}
Physical Distances \\
Connectivity through Edge Matrix \\
Connectivity through Edge Matrix \\
Brain Region Connectivity \\
Physical Distances
\end{tabular} \\ \hline

Distance &
\begin{tabular}[c]{@{}c@{}}
EGFG \cite{egfg}\\
JSCFE \cite{jscfe}\\
MST-GNN \cite{mst-gnn}\\
TARDGCN \cite{tardgcn} \\
OMHGL \cite{omhgl}
\end{tabular} &
\begin{tabular}[c]{@{}c@{}}
Euclidean distance \\
Euclidean distance \\
Phase Lag Index \\
Cosine distance \\
Cosine distance 
\end{tabular} \\ \hline

Similarity &
\begin{tabular}[c]{@{}c@{}}
MD-AGCN \cite{md-agcn}\\
HetEmotionNet \cite{hetemotionnet}\\
ST-GCLSTM \cite{st-gclstm}\\
HD-GCN \cite{hd-gcn}\\
HB-SR \cite{hb-sr}
\end{tabular} &
\begin{tabular}[c]{@{}c@{}}
Pearson Correlation Coefficient \\
Mutual Information \\
Pearson Correlation Coefficient \\
Gaussian Similarity \\
Phase Locking Value
\end{tabular} \\ \hline \hline
\end{tabular}%
}
\label{tb: model-inde}
\end{table}

\subsection{Model-dependent edge}

Model-dependent edge refers to the computation of edges using model parameters. The edges containing model parameters can be dynamically updated with training so that the model can be generalized over different data. 
Compared to Model-independent edges that are limited by prior knowledge or vector relationships, Model-dependent edge can capture more complex brain region relationships in emotional EEG without manually adjusting when handling different signals.
However, additional parameters make it challenging to balance accuracy and computational burden.
Model-dependent edge contains four categories, as shown in Table \ref{tb: para}.

\noindent \textbf{Parametric} edge refers to directly using the model parameter matrix as the edge matrix. As the most direct method of using model parameters, Parametric edge is the most commonly used edge of Model-dependent edge. Parametric edge is updated as follows:

{\normalsize
\begin{equation}
\mathbf{E}'_{i,j}=(1-\rho) \mathbf{E}_{i,j}+\rho \frac{\partial \mathcal{L}}{\partial \mathbf{E}_{i,j}}
\end{equation}
}where $\rho$ denotes the learning rate. $\frac{\partial \mathcal{L}}{\partial \mathbf{E}}$ refers to the partial derivative of the loss $\mathcal{L}$ with respect to $\mathbf{E}_{i,j}$.

\noindent \textbf{Weighted} edge indicates that the edge matrix can be viewed as the product of the node relationships and the model parameters used as weights.
Therefore, the essential meaning of Weighted edge depends on the node relationships.
Weighted edge can introduce prior knowledge or vector calculations to constrain the edge by setting it as the node relationships. The weighted edge implementation of existing methods is as follows:

{\normalsize
\begin{equation}
\mathbf{E}_{i,j}=  \xi (\mathbf{W} * f[v_i,v_j])
\end{equation}
}where $\mathbf{W}$ represents the model parameter utilized as weight. $\xi(\cdot)$ is activation function. $f[\cdot,\cdot]$ denotes the relationship function between nodes. Siam-GCAN \cite{siam-gcan} and DBGC-ATFFNet-AFTL \cite{dbgc-atffnet-aftl} utilize distance between nodes as the relationship function. LR-GCN \cite{lr-gcn} and STFCGAT \cite{stfcgat} concatenate any two nodes to perform a linear transformation as the relationship function.

\noindent \textbf{Bias} edge adding model parameters as bias terms to the node relationships as the edge matrix. It differs from Weighted edge only in the meaning of the parameters. The implementation of Bias edge is as follows:

{\normalsize
\begin{equation}
\mathbf{E}_{i,j} =  \mathbf{W} + f[v_i,v_j]
\end{equation}
}where $\mathbf{W}$ represents the model parameter utilized as bias. MRGCN \cite{mrgcn} utilizes the calculation of the physical distance between nodes as the relationship function. OGSSL \cite{ogssl} applies the similarity between nodes as the relationship function.

\noindent \textbf{Subspace} edge denotes the projection of nodes to subspace by parameters to obtain similarity as edges.
It is similar in principle to the Attention mechanism and allows the model to capture the complex relationship between every two nodes. However, it adds more model complexity and computational cost than other Model-dependent edges. MD-AGCN \cite{md-agcn}, JSCFE \cite{jscfe}, HGCN \cite{zhao2025heterogeneous}, VBH-GNN \cite{liu2024vbh} and ASTG-LSTM \cite{astg-lstm} use the following to implement Subspace edge:

{\normalsize
\begin{equation}
E_{i,j}=\xi (v_i^T*\mathbf{W}_1^T*\mathbf{W}_2*v_j)
\end{equation}
}$\mathbf{W}_1$ and $\mathbf{W}_2$ are projection matrices for two nodes.

\begin{table}[t]
\caption{Overview of Model-dependent edge.}
\vspace{-10pt}
\centering
\renewcommand{\arraystretch}{1} 
\resizebox{0.85\columnwidth}{!}
{
\scriptsize
\begin{tabular}{c|c|c}
\hline \hline
\multicolumn{1}{c|}{\rule[-1ex]{0pt}{3.5ex}\footnotesize \textbf{Edge}} &
\multicolumn{1}{c|}{\footnotesize \textbf{Baseline}} &
\multicolumn{1}{c}{\footnotesize \textbf{Edge Feature}} \\ \hline \hline

Parametric &
\begin{tabular}[c]{@{}c@{}}
DGCNN \cite{dgcnn} \\
GCB-Net \cite{gcb-net} \\
HD-GCN \cite{hd-gcn} \\
AHGCN \cite{ahgcn} \\
GIGN \cite{gign} \\
EmoGT \cite{emogt} \\
EmT \cite{ding2025emt} \\
DTC-GCN \cite{zhou2025eeg} \\
RPGCN \cite{zhou2025rpgcn} \\
VSGT \cite{liu2024vsgt}
\end{tabular} &
\begin{tabular}[c]{@{}c@{}}
Gaussian similarity \\
Learned Topology \\
Gaussian similarity \\
Learned Topology \\
Learned Topology \\
Learned Topology \\
Multi-view Learning \\
Sparse Causality \\
Relational Probabilistic \\
Prior Constrained Dependencies
\end{tabular} \\ \hline

Weight &
\begin{tabular}[c]{@{}c@{}}
Siam-GCAN \cite{siam-gcan}\\
DBGC-ATFFNet-AFTL \cite{dbgc-atffnet-aftl} \\
STFCGAT \cite{stfcgat} \\
LR-GCN \cite{lr-gcn}
\end{tabular} &
\begin{tabular}[c]{@{}c@{}}
Distance-based Weights \\
Distance-based Weights \\
Linear Transformation \\
Attention Weights \& Linear Transformation
\end{tabular} \\ \hline

Bias &
\begin{tabular}[c]{@{}c@{}}
OGSSL \cite{ogssl} \\
MRGCN \cite{mrgcn}
\end{tabular} &
\begin{tabular}[c]{@{}c@{}}
Euclidean Similarity \\
Physical Distance
\end{tabular} \\ \hline

Subspace &
\begin{tabular}[c]{@{}c@{}}
MD-AGCN \cite{md-agcn} \\
ASTG-LSTM \cite{astg-lstm}\\
HGCN \cite{zhao2025heterogeneous} \\
VBH-GNN \cite{liu2024vbh}

\end{tabular} &
\begin{tabular}[c]{@{}c@{}}
Pearson Correlation Coefficient \\
Attention Mechanism \\
Heterogeneous Multi-head \\
Variational Bayesian Graph Inference 

\end{tabular} \\ \hline \hline

\end{tabular}%
}
\label{tb: para}
\end{table}

\section{Graph-level Stage}
\label{sec: graph}

There are four graph structures included in the Graph-level Stage: Multi-graph, Hierarchical graph,  Time series graph, and Sparse graph. According to the dependencies that graphs indicate, Multi-graph is divided into Horizontal\&Vertical graph, Temporal\&Frequency graph, and Local\&Global graph. Hierarchical graph is divided into Dynamic hierarchy graph and Predetermined hierarchy graph. 
The realisation of Time series graph is divided into Temporal graph and Temporal encoder. The sparsity implementation of Sparse graph consists of Threshold and Sparse weight.

\subsection{Multi-graph}
\label{sec: multi}

Multi-graph allows the model to capture different emotional EEG dependencies simultaneously by concatenating multiple types of graph embeddings. Various relational dependencies exist in emotional EEG, such as temporal, frequency, and local brain region dependency, etc. The multi-graph structure can learn and consider different dependencies simultaneously, which helps to build relationships in emotional EEG more comprehensively. However, the multi-stream graph modeling process increases the computational burden. The existing Multi-graph can be subdivided into three categories, as shown in Table \ref{tb: multi}.

\noindent \textbf{Horizontal\&Vertical} graph captures both vertical and horizontal spatial dependencies of brain regions. There are differences in the horizontal and vertical connectivity patterns of brain regions so that the Horizontal\&Vertical structure can model more physiologically plausible spatial dependency. ERHGCN \cite{erhgcn} and BiHDM \cite{bihdm} both utilize this graph. BiHDM further constructed different horizontal and vertical graphs in the left and right hemispheres based on the existence of imbalance between hemispheres.

\noindent \textbf{Temporal\&Frequency} graph combines the temporal dependency and frequency dependency of the emotional EEG signal. 
Temporal dependence allows the model to notice amplitude features that are highly correlated with emotion. Frequency dependence encompasses activating the corresponding frequency band in a particular emotion, such as the increase in the $\alpha$ band during the positive emotion. Existing methods MD-AGCN \cite{md-agcn}, HetEmotionNet \cite{hetemotionnet}, and DBGC-ATFFNet-AFTL \cite{dbgc-atffnet-aftl} all adopt a similar two-stream structure, where each stream corresponds to the temporal-spatial and frequency-spatial space, and finally graph embeddings are concatenated and fed into the classifier.

\noindent \textbf{Local\&Global} graph combines global dependency between brain regions and local dependency within brain regions. In emotional states, global dependence responds to interactions between brain regions, such as activity in surrounding brain regions caused by activity in specific brain regions. Local dependence focuses on relationships within specific brain regions, e.g., activity within the frontal lobe and cortex is associated with positive emotions. GECNN \cite{gecnn}, HD-GCN \cite{hd-gcn}, AHGCN \cite{ahgcn}, and LGGNet \cite{lggnet} all infer the spatial dependencies of all channels and channels within specific brain regions. Similarly, MRGCN \cite{mrgcn} introduces short-range and long-range spatial dependencies corresponding to localized intra-region correlations and inter-region correlations.

\begin{table}[t]
\caption{Overview of Multi-graph.}
\vspace{-10pt}
\centering
\renewcommand{\arraystretch}{1}
\resizebox{0.85\columnwidth}{!}
{
\scriptsize
\begin{tabular}{c|c|c}
\hline \hline
\multicolumn{1}{c|}{\rule[-1ex]{0pt}{3.5ex}\footnotesize \textbf{Graph}} &
\multicolumn{1}{c|}{\footnotesize \textbf{Baseline}} &
\multicolumn{1}{c}{\footnotesize \textbf{Graph Details}} \\ \hline \hline

Horizontal\&Vertical &
\begin{tabular}[c]{@{}c@{}}
BiHDM \cite{bihdm}\\
ERHGCN \cite{erhgcn}
\end{tabular} &
\begin{tabular}[c]{@{}c@{}}
Bi-hemispheric Paired Discrepancy \\
Brain Regions \\
\end{tabular} \\ \hline

Temporal\&Frequency &
\begin{tabular}[c]{@{}c@{}}
MD-AGCN \cite{md-agcn}\\
HetEmotionNet \cite{hetemotionnet}\\
DBGC-ATFFNet-AFTL \cite{dbgc-atffnet-aftl} \\
VBH-GNN \cite{liu2024vbh}
\end{tabular} &
\begin{tabular}[c]{@{}c@{}}
Two-stream Structure \\
Heterogeneous Domain Fusion \\
Two-stream Structure \\
Two-stream Structure
\end{tabular} \\ \hline

Local\&Global &
\begin{tabular}[c]{@{}c@{}}
GECNN \cite{gecnn}\\
HD-GCN \cite{hd-gcn}\\
AHGCN \cite{ahgcn}\\
MRGCN \cite{mrgcn}\\
LGGNet \cite{lggnet}
\end{tabular} &
\begin{tabular}[c]{@{}c@{}}
Electrodes Groups \\
Brain Region \\
Brain Region \\
Long/Short Range Dependencies\\
Brain Region
\end{tabular} \\ \hline \hline
\end{tabular}%
}
\label{tb: multi}
\end{table}

\subsection{Hierarchical graph}

Hierarchical graph groups channels to allow the model to capture local spatial dependency within specific brain regions. The spatial dependency of emotional EEG consists of inter-region and intra-region dependencies. The former corresponds to the region-level synergy between different brain regions under emotion, and the latter corresponds to the neural-level activity within the brain region associated with a specific emotion, e.g., the frontal cortex and lobe are related to positive emotions. 
Therefore, the hierarchical graph structure allows the model to effectively utilize the signal features contained in these local brain regions that are directly related to emotions. 
The difference from the Local\&Global multi-graph structure is that the hierarchical graph structure does not necessarily employ a multi-stream process, thus reducing the computational burden. The existing Hierarchical graph contains two categories, as shown in Table \ref{tb: hier}.

\noindent \textbf{Dynamic hierarchy} graph indicates that the model adaptively groups channels and then infers spatial dependency between groups. There are significant individual differences in emotional EEG between subjects, which implies that different subjects will have subtle differences in brain region correlations during the same emotion. Therefore, dynamic hierarchy can reduce the effect of individual differences. AHGCN \cite{ahgcn}, SCC-MPGCN \cite{scc-mpgcn}, and MDGCN-SRCNN \cite{mdgcn-srcnn} firstly infer channel-level spatial dependency, then implement a trainable weight assignment matrix to assign different electrodes to different brain regions, and finally build region-level spatial graph. MST-GNN \cite{mst-gnn} also implements channel-level dependency first, and then, based on the minimum spanning tree, it starts to search for children nodes with the strongest correlation so that each branch can be regarded as a group.

\noindent \textbf{Predetermined hierarchy} graph denotes the setting of the brain regions to which the electrodes belong based on existing prior knowledge. 
Prior knowledge refers to the cytoarchitecture of the brain, i.e., the division of brain regions according to the function. Predetermined hierarchy graph is physiologically reasonable and does not introduce additional computational burden. GMSS \cite{gmss}, R2G-STNN \cite{r2g-stnn}, and VPR \cite{vpr} apply a local-to-global process. They divide electrodes into multiple regions in advance based on prior knowledge and first infer spatial dependency within regions and then between regions. In contrast, V-IAG \cite{v-iag} utilizes a global-to-local process. It first infers fully connected spatial dependency and then divides electrode regions and aggregates region-level nodes to construct relationships between regions. HD-GCN \cite{hd-gcn}, and LGGNet \cite{lggnet} adopt the multi-graph structure in Section \ref{sec: multi}, where local and global dependencies are reasoned in parallel and then concatenated together.

\begin{table}[t]
\caption{Overview of Hierarchical graph.}
\vspace{-10pt}
\centering
\renewcommand{\arraystretch}{1}
\resizebox{0.85\columnwidth}{!}
{
\scriptsize
\begin{tabular}{c|c|c}
\hline \hline
\multicolumn{1}{c|}{\rule[-1ex]{0pt}{3.5ex}\footnotesize \textbf{Graph Type}} &
\multicolumn{1}{c|}{\footnotesize \textbf{Baseline}} &
\multicolumn{1}{c}{\footnotesize \textbf{Hierarchy Details}} \\ \hline \hline

Dynamic hierarchy &
\begin{tabular}[c]{@{}c@{}}
AHGCN \cite{ahgcn}\\
SCC-MPGCN \cite{scc-mpgcn} \\
MST-GNN \cite{mst-gnn} \\
MDGCN-SRCNN \cite{mdgcn-srcnn}
\end{tabular} &
\begin{tabular}[c]{@{}c@{}}
Channels/Regions Spatial Dependency\\
Channels/Regions Spatial Dependency \\
Channels/Regions Spatial Dependency \\
Correlation-based Tree Structure
\end{tabular} \\ \hline

Predetermined hierarchy &
\begin{tabular}[c]{@{}c@{}}
R2G-STNN \cite{r2g-stnn} \\
VPR \cite{vpr} \\
V-IAG \cite{v-iag}\\
HD-GCN \cite{hd-gcn}\\
GMSS \cite{gmss} \\
LGGNet \cite{lggnet}
\end{tabular} &
\begin{tabular}[c]{@{}c@{}}
Local to Global \\
Local to Global \\
Global to Local \\
Multi-Graph Concatenation \\
Local to Global \\
Multi-Graph Concatenation 
\end{tabular} \\ \hline \hline
\end{tabular}%
}
\label{tb: hier}
\end{table}

\subsection{Time series graph}
\label{sec: time}

Time series graph indicates that the model decomposes the signal into multiple time slices and constructs temporal dependency between slices. 
The temporal dependency of emotional EEG directly reflects the relationship between signal amplitude change and emotion. 
A high degree of activation within 300ms in the frontal cortex channels is usually associated with positive emotions, while a low degree of activation is usually associated with negative emotions. Therefore, Time series graph enables the model to construct and exploit the spatial and temporal dependencies of emotional EEG. 
Typically, it is implemented by splitting the input signal into multiple time slices to model spatial graphs within slices and then deriving the temporal dependency between slices by temporal graph or temporal encoder, as shown in Table \ref{tb: time}.

\noindent \textbf{Temporal graph} represents the model regards spatial graph embeddings of time slices as nodes to construct a temporal graph.
CR-GAT \cite{cr-gat}, VSGT \cite{liu2024vsgt} and GIGN \cite{gign} employ a GNN in the temporal dimension to construct a temporal graph by regarding time slices as nodes. MD-AGCN \cite{md-agcn} couples the spatial graphs of all time slices and then takes the average as a temporal graph.

\noindent \textbf{Temporal encoder} represents that the model combines GNN and temporal encoder to derive the spatial and temporal dependencies of the emotional EEG. This derivation process is usually divided into two steps; firstly, the GNN is used to infer the spatial dependency within the time slices and update the spatial graph embeddings; then, the graph embedding of all the time slices is concatenated together and fed into a temporal encoder. ST-GCLSTM \cite{st-gclstm}, ASTG-LSTM \cite{astg-lstm}, R2G-STNN \cite{r2g-stnn} and DTC-GCN \cite{zhou2025eeg} use LSTM as the temporal encoder. Similarly, HetEmotionNet \cite{hetemotionnet} uses GRUs, and EmoGT \cite{emogt} and EmT \cite{ding2025emt} utilizes a transformer encoder.

\begin{table}[t]
\caption{Overview of Time series graph.}
\vspace{-10pt}
\centering
\renewcommand{\arraystretch}{1.0}
\setlength{\tabcolsep}{15pt}
\resizebox{0.95\columnwidth}{!}
{
\scriptsize
\begin{tabular}{c|c|c}
\hline \hline
\multicolumn{1}{c|}{\rule[-1ex]{0pt}{3.5ex}\footnotesize \textbf{Methods}} &
\multicolumn{1}{c|}{\footnotesize \textbf{Baseline}} &
\multicolumn{1}{c}{\footnotesize \textbf{Temporal Details}} \\ \hline \hline

Temporal Graph &
\begin{tabular}[c]{@{}c@{}}
MD-AGCN \cite{md-agcn} \\
CR-GAT \cite{cr-gat}\\
GIGN \cite{gign} \\
VSGT \cite{liu2024vsgt}
\end{tabular} &
\begin{tabular}[c]{@{}c@{}}
Average Spatial Graphs \\
Time Slices as Nodes \\
Time Slices as Nodes \\
Time Slices as Nodes 
\end{tabular} \\ \hline

Temporal Encoder &
\begin{tabular}[c]{@{}c@{}}
R2G-STNN \cite{r2g-stnn}\\
ASTG-LSTM \cite{astg-lstm}\\
HetEmotionNet \cite{hetemotionnet}\\
ST-GCLSTM \cite{st-gclstm} \\
EmoGT \cite{emogt} \\
EmT \cite{ding2025emt}\\
DTC-GCN \cite{zhou2025eeg}
\end{tabular} &
\begin{tabular}[c]{@{}c@{}}
LSTM \\
LSTM \\
GRUs \\
LSTM \\
Transformers \\
Transformers \\
LSTM
\end{tabular} \\ \hline \hline
\end{tabular}%
}
\label{tb: time}
\end{table}

\subsection{Sparse graph}

Sparse graph implies that the graph connectivity relations are sparse. 
There is a concentration of brain activity in areas directly related to emotion, which means that most connections are weakly associated with emotion. 
Therefore, specific emotions can be identified by the electrical signals that are active around specific brain regions, such as the frontal cortex and lobe.
The sparse graph structure reduces the influence of irrelevant information and retains only the essential connections related to a specific emotion. 
It differs from the hierarchical structure in that it couples specific groups of channels to achieve fewer connections, whereas the sparse graph structure retains fewer edges by filtering directly on top of the fully connected graph. 
Existing methods set \textbf{Threshold} or add \textbf{Sparse weights} to the loss function to retain a specific number of edges to form a sparse graph. 
DAGAM \cite{dagam} and SOGNN \cite{sognn} set Threshold $k$ to achieve sparsity. DAGAM proposes self-attention graph pooling that regards the Score matrix as the sparse weight matrix and retains top-$k$ elements to retain top score edges. SOGNN utilizes a $1 \times 2$ max pooling layer to retain top-$k$ edges.
SGA-LSTM \cite{sga-lstm}, and SparseDGCNN \cite{sparsedgcnn} set Sparse weight to measure the contributions of different channels. 
The sparse weight is added to the loss to update the edge matrix. The sparsity is achieved when some elements of the edge matrix are close to 0.

\section{Future Directions}

Despite the progress made by existing methods, there are still some challenges and possible future directions. We analyze several directions that are significant for future EEG-based emotion recognition.

\noindent \textbf{Temporal fully-connected graph} represents a fully-connected temporal graph across time slices. It aims to solve the problem of incomplete temporal dependency in current EEG-based emotion recognition.
The temporal dependency of emotional EEG inferred by existing GNNs lacks the correlation between different channels across time slices.
For Time series graph mentioned in Section \ref{sec: time}, although the relationship between time slices can be established by the Temporal graph or encoder, this relationship is limited between the same electrodes across time slices, such as $v_i^{t-1}$ and $v_i^{ t}$, while relations between different electrodes across time slice are ignored, such as $v_i^{t-1}$ and $v_j^{t}$. 
This hetero-electrode relationship across time slice corresponds to a delayed response in the emotional state of the brain, i.e., there is asynchrony in the interaction between brain regions. 
The existence of delays between specific brain regions and surrounding brain regions for brain region activity elicited by emotional excitation implies that this asynchronous relationship contains information directly related to emotion.
Therefore, a more complete temporal dependency constructed by the Temporal fully-connected graph is expected to improve the accuracy of emotion recognition.

\noindent \textbf{Graph condensation} aims to compress a large and complex graph structure into a simple and compact representation and retain essential emotion-related information. 
Both coarsening and sparsifying are promising techniques for future EEG-based emotion recognition. Emotional EEG tends to move towards high resolution.
Current GNNs in EEG-based emotion recognition utilize fully connected graphs, where nodes usually have an excessive number of neighbors, which provides redundant information and even noise.
As the number of electrodes increases, this defect becomes more significant. Therefore, graph condensation techniques can not only improve training efficiency but also weaken the influence of emotionally irrelevant information.
The activity of brain regions in emotional states is focused, which implies that the information of specific brain regions is sufficient to identify specific emotions, while most of the connections are emotionally irrelevant. In practice, GNNs in EEG-based emotion recognition are trained to preserve localization.
Existing GNNs in this field generally need to run the training process at least once to find the optimal graph condensation, so if the optimal graph is identified earlier, e.g., by training for only a few epochs, then this will significantly accelerate the running process of GNNs to build emotional representation.

\noindent \textbf{Heterogeneous graph} can compose different types of entities (i.e., different physiological signals). It can not only provide the graph structure of the data associations but also provide higher-level semantics of the physiological signals.
Most of the current EEG-based emotion recognition datasets contain multiple physiological signals, while the heterogeneous graph is rarely used.
Human emotion regulation includes multiple physiological systems, such as the cardiovascular and exocrine systems. 
Therefore, emotional dependencies exist not only in the interactions between brain regions but also in the activities of multiple physiological systems (organs) regulated by the brain. 
For example, the correlation between the heart and the striatum region increases when positive emotions are generated; the correlation between the heart and the right prefrontal cortex increases when negative emotions are generated.
Heterogeneous graphs can expand the emotional dependencies of specific brain regions with other physiological organs based on the construction of emotional dependencies within the brain.
In other words, Heterogeneous graphs benefit from the ability to construct more comprehensive dependencies and become a promising direction for EEG-based emotion recognition.

\noindent \textbf{Dynamic graph} refers to a graph whose structure changes dynamically over time. It differs from dynamically updated GNNs in EEG-based emotion recognition, meaning the graph structure can be updated automatically as training proceeds. 
This belongs to model-level dynamic, while the dynamic graph is graph structure-level dynamic.
Given dynamic graph $\textbf{G}(\textbf{V},\textbf{E})$,  $\textbf{V} = \{(v, t_s, t_e)\}$ and $\textbf{E} = \{(e, t_s, t_e)\}$. $t_s$ and $t_e$ are timestamps when edges and nodes appear and disappear.
Existing methods reason about the temporal dependency of emotional EEG by cropping time slices and constructing spatial graphs independently. 
The former indicates that the graph can be updated, and the latter indicates that what the graph represents is the dynamic temporal dependency in emotional EEG.
Although temporal graphs or encoders can be used to build temporal dependencies between slices, such dependencies do not incorporate hetero-electrode relations across time slices, i.e., changes of edges across time. 
In contrast, the dynamic dependency depicted by the dynamic graphs encompasses changes in the edges and, therefore, corresponds to the complete time dependency of the emotional EEG.

\section{Conclusion}

Graph neural networks have greatly facilitated the development of EEG-based emotion recognition. This survey provides a comprehensive study of existing GNNs in this field. 
In-depth discussions and summaries of the reviewed methods are presented, as well as categorization according to the three stages of graph construction.
Future directions for GNN design addressing the existing challenges are also proposed. 
We hope that this survey will provide clear guidance for building GNNs in the field of EEG-based emotion recognition.

\subsubsection{\discintname}
The authors have no competing interests to declare that are relevant to the content of this article.

\begin{credits}

\subsubsection{\ackname} This study was supported in part by the Hubei Provincial Natural Science Foundation of China under Grant 2024AFB932, and in part  by the CCF-NSFOCUS 'Kunpeng' Research Fund.

\end{credits}

\clearpage
\bibliographystyle{splncs04}
\bibliography{references}
%





\end{document}